\documentclass[aps,twocolumn,pra,superscriptaddress,amsmath,showpacs,tightenlines]{revtex4-1}
\usepackage{amssymb}
\usepackage{amsmath}
\usepackage{graphicx}
\usepackage{subfigure}
\usepackage{natbib}
\usepackage{epsfig}
\usepackage{amsfonts}
\usepackage{mathrsfs}
\usepackage{xcolor}
\usepackage[toc,page,title,titletoc,header]{appendix}
\begin{document}

\title{Photonic state engineering via energy-level crossing by giant atoms in topological waveguide QED setup}

\author{Mingzhu Weng}
\affiliation{School of Physics Science and Technology, Shenyang Normal University, Shenyang 110034, China}
\author{Gang Wang}
\email{gangwang82@suda.edu.cn}
\affiliation{School of Physical Science and Technology, Soochow University, Suzhou 215006, China}
\author{Zhihai Wang}
\email{wangzh761@nenu.edu.cn}
\affiliation{Center for Quantum Sciences and School of Physics, Northeast Normal University, Changchun 130024, China}

\begin{abstract}
Photonic state engineering in waveguide QED is typically based on local light-matter interactions. This limits its control over the spatial structure of bound photonic states. Here, we demonstrate a distinct mechanism arising from the interplay between nonlocal giant-atom coupling and topological band structure. Specifically, we consider giant atoms coupled to a Su-Schrieffer-Heeger waveguide and show that this configuration enables a controllable energy-level crossing protected by the topological gap. Adiabatically sweeping the atomic detuning across the crossing leads to a controlled exchange between distinct photonic bound states. In a two-giant-atom configuration, this mechanism achieves high-fidelity conversion of a spatially splitting state into a combining state. Extending this scheme to three-giant atoms, we further realize robust, shape-preserving photon transfer mediated by sequential in-gap crossings. Our results demonstrate how topology and nonlocal light-matter coupling can be combined to achieve programmable control of bound photonic states in waveguide QED platforms.
\end{abstract}


\maketitle
\section{introduction}
Waveguide systems play a fundamental role in controlling light propagation and light-matter interactions, driving advances in topological photonics and quantum information technologies~\cite{DR2017,CN2017,TO2019}. In recent years, topological photonics has emerged as a powerful framework for designing novel optical behaviors, drawing inspiration from topological phases in condensed matter physics~\cite{ZL2022,LL2014,AB2017,MZ2010,JS2019}. Through tailored designs in photonic crystals, metamaterials, and coupled resonator arrays, researchers have demonstrated remarkable phenomena including robust unidirectional transport and disorder-resistant edge states~\cite{ZW2009,PS2017,MR2018,YO2020,MH2011,WJ2014}. These advances pave the way for harnessing topological protection to enhance the coherence and operational robustness of photonic devices~\cite{MS2019,YW2017,SB2018}. This has prompted a shift in research focus from using waveguides as data bus to transmitting atomic states. In waveguides, photons can be modulated during transmission. Therefore, the transmission of photon states with specific shape distributions should also be given attention. A typical model for this topologically protected transmission of photon states is the SSH model~\cite{WP1979}, which serves as a platform for exploring light propagation and topological states within photonic environments~\cite{MB2019,EK2021,AB2016,WC2019,LL2016,ZK2023}.

The SSH model has garnered significant attention due to its potential for implementing photonic state transfer, primarily because of its topologically protected edge states, which facilitate robust, unidirectional propagation that is immune to disorder~\cite{TT2022,NE2021,CW2022,WL2022,LH2022,FM2018,YC2023}. Meanwhile, superconducting quantum circuits have proven to be a versatile platform for simulating topological physics~\cite{XG2017,PR2014,EF2017}. Recent experiments have further demonstrated that a single superconducting qubit can realize topological phase transitions and reveal associated band structures~\cite{MD2014,ZY2020,XS2018,XS2019}. When atoms are coupled to an SSH chain, distinctive quantum optical phenomena emerge, such as chiral interactions that give rise to topologically protected light transport and edge states~\cite{XLD2022,WN2020,WN2020lett,YZ2022,LQ2021}. By engineering atomic interactions, photon transfer can be effectively modulated~\cite{XY2025,LZ2022}. Therefore, coupling atoms to a waveguide facilitates precise control over the photon transfer process, supporting scalable quantum information processing~\cite{LL2021,QC2015,JS2015}.

Unlike natural atoms, giant atoms composed of superconducting qubits enable nonlocal coupling with waveguides~\cite{MV2014,RM2017,AM2021}, leading to distinct interference and time delay effects~\cite{XL2023,AC2023,BK2020,MW2025,DW2024} and facilitate phenomena like frequency-dependent atomic relaxation rates~\cite{YT2022,XL2022,MW2024}, chiral light-matter interactions~\cite{AF2018,AC2020,XW2022}, as well as non-Markovian dynamics~\cite{GA2019,LD2023,XY2022,SG2020,XJ2023}. The underlying physics of these phenomena arises from the interference of photons at multiple spatial coupling points of a single giant atom to the waveguide. Consequently, the nonlocal coupling of giant atoms allows for advanced manipulation of photons, including single-photon scattering~\cite{WZ2020,WZ2024} and chiral radiation~\cite{CJ2023}. Motivated by these distinctive effects, we explore in this work the potential of giant atoms to dynamically engineer the spatial distribution of photons within a topological waveguide.

In this paper, we investigate the system of giant atoms coupled to the SSH chain. We find that an energy-level crossing structure can be engineered within the band gap by adjusting the geometric configuration of the giant atoms, which is not possible in waveguide without topology. Based on the engineered energy-level crossing structure, we achieve control over photonic states. Specifically, in the two giant atom system, it enables dynamic manipulation of the photonic state shape, allowing conversion from a splitting state to a combining state. In the three giant atom system, the same mechanism facilitates shape-preserving transfer. Our work leverages the nonlocal coupling of giant atoms to propose a strategy for photon manipulation in topological systems.

The rest of the paper is organized as follows. In Sec.~\ref{Model}, we introduce the theoretical model and analyze the energy spectrum of the system. The band gap states are further classified according to their photonic distribution characteristics. In Sec.~\ref{evolution}, we discussed the manipulation scheme for the shape of photon from splitting state to combining state. In Sec.~\ref{discussion}, we presented a shape-preserving transfer scheme for photon states in three giant atomic configuration. In Sec.~\ref{conclusion}, we provide a brief summary and discussion.

\section{Model and Energy Spectrum}
\label{Model}
\begin{figure}
\centering
\includegraphics[width=1\columnwidth]{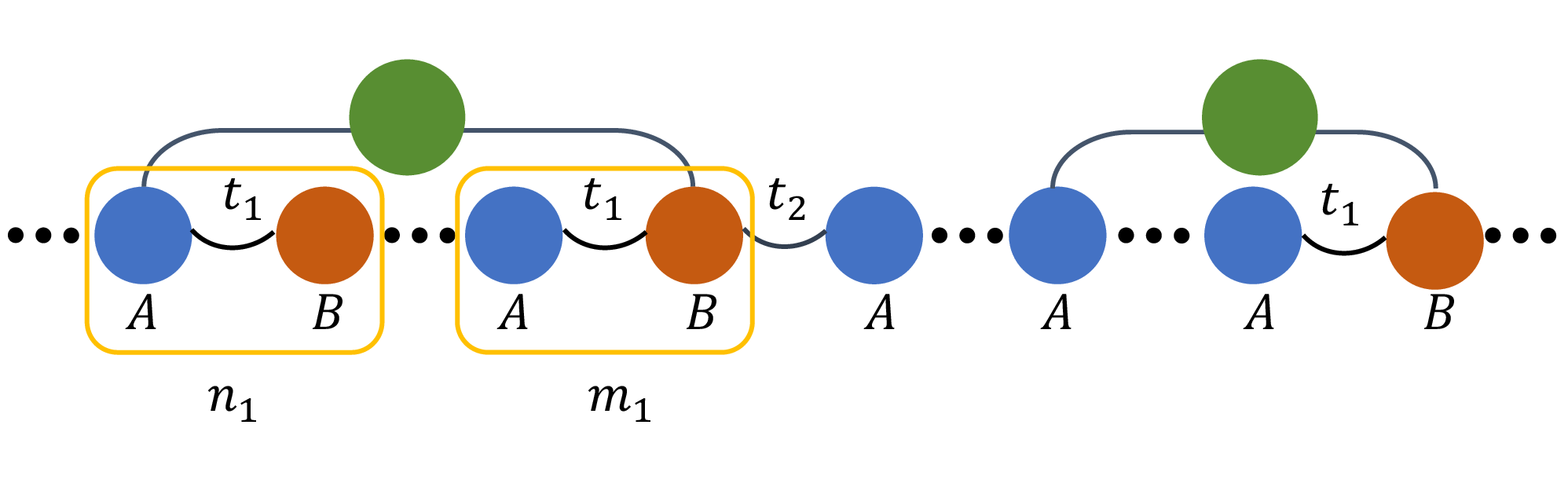}
\caption{Schematic configuration of two giant atoms coupled to the SSH chain via $A-B$ coupling. The green balls represent giant atoms coupled to the waveguide via two sites in the separate configuration. The left leg of the $i$th giant atom couples to the sublattice $A$ of the $n_{i}$th cell and the right leg couple to the sublattice $B$ of the $m_{i}$th cell. The blue balls denote sublattice $A$ of the SSH chain and the orange balls denote sublattice $B$. The intracell and intercell hopping strengths are $t_{1}$ and $t_{2}$, respectively.}
\label{device}
\end{figure}

\begin{figure}
\centering
\includegraphics[width=1\columnwidth]{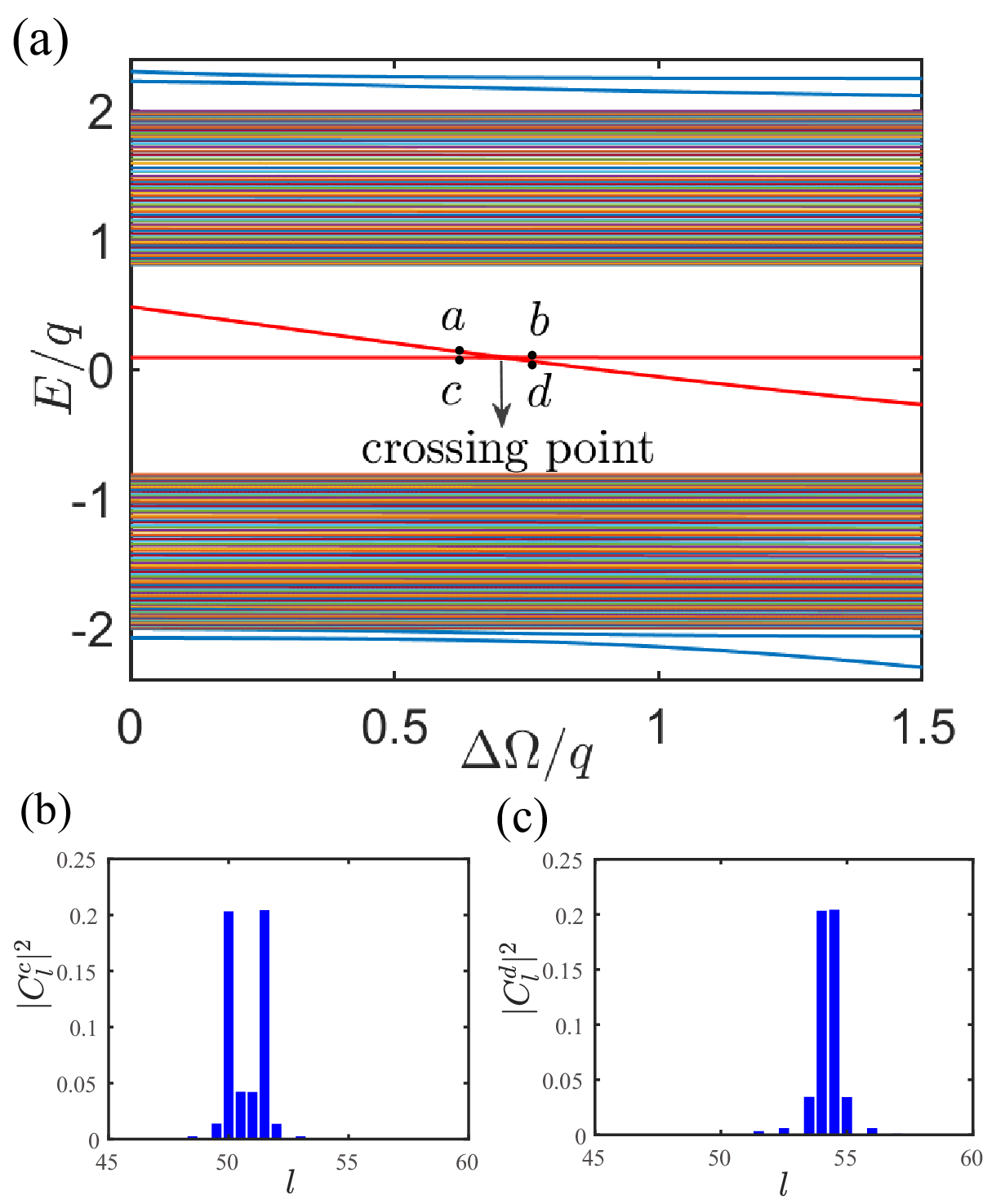}
\caption{(a) The energy spectrum of the system varies with the frequency detuning $\Delta\Omega$. The red solid lines denote the two band gap states. The black points on the left side of the crossing point correspond to the frequency detuning $\Delta\Omega=0.65q$. The parameters selected for the black points on the right side is $\Delta\Omega=0.75q$. The photonic probability distribution of the lower band gap states as the function of the position for (b) $|\psi_{c}\rangle$ with point $c$ and (c) $|\psi_{d}\rangle$ with point $d$, respectively. The other parameter set as $\theta=0.2\pi, L=100, n_{1}=50, m_{1}=52, n_{2}=54, m_{2}=55, \omega_{1}=0.4q, \delta=0.5q,$ and $g=0.9q$. }
\label{energytwo}
\end{figure}

As schematically shown in Fig.~\ref{device}, we consider multiple giant atoms in the system, each of which couples to an SSH chain waveguide via two points. The SSH chain waveguide is described by two resonators $A/B$ of size $L$ with alternating nearest-neighbor hoppings $t_1$ and $t_2$. Under periodic boundary, the Hamiltonian of the SSH chain is
\begin{eqnarray}
H_{\rm SSH}&=&\sum_{l=1}^{L}\omega(c_{A,l}^{\dagger}c_{A,l}+c_{B,l}^{\dagger}c_{B,l})\nonumber\\
&&+(t_{1}c_{A,l}^{\dagger}c_{B,l}+t_{2}c_{A,l+1}^{\dagger}c_{B,l}+{\rm H.c.}).
\end{eqnarray}
As shown by the orange box in the Fig.~\ref{device}, we assume that each unit cell hosts two resonators $A$ and $B$ with the identical frequency $\omega$, which we now set to zero, i.e., $\omega=0$.  $c_{A(B),l}^{\dagger}(c_{A(B,l)})$ are the creation (annihilation) operators of the $A/B$ resonators in $l {\rm th}$ unit cell. The intracell and intercell couplings are $t_1=q(1+\delta\cos\theta)$ and $t_2=q(1-\delta\cos\theta)$ with $\delta$ being the dimerization strength. Applying Fourier transform $c_{A(B),k}=1/\sqrt{L}\sum_{k}e^{ikl}c_{A(B),l}$ and defining $V^{\dagger}=(c_{A,k}^{\dagger},c_{B,k}^{\dagger})$, the SSH Hamiltonian can be written in momentum space as $H_{\rm SSH}=\sum_{k}V^{\dagger}\tilde{H}_{\rm SSH}(k)V$ with
\begin{eqnarray}
\tilde{H}_{\rm SSH}=\begin{pmatrix} 0 & t_{1}+t_{2}e^{-ik} \\ t_{1}+t_{2}e^{ik} & 0 \end{pmatrix}.
\end{eqnarray}
The dispersion relation of the SSH model can be calculated as $E_{k\pm}=\pm \omega_{k}=\pm \sqrt{t_{1}^{2}+t_{2}^{2}+2t_{1}t_{2}\cos(k)}$. Therefore, the SSH model exhibits a twofold degenerate two-band structure with  $\omega_k=\omega_{-k}$.

The $i$th giant atom are designed to couple with the SSH chain at cells $m_{i}$ and $n_{i}$ as shown in Fig.~\ref{device}. The interaction Hamiltonian between the giant atom and the topological waveguide for the $A-B$ coupling is given as
\begin{equation}
H_{I}=\sum_{i=1}^{N}\omega_i(t)|e\rangle\langle e|+g_{i}\sigma_{i}^{+}(c_{A,n_{i}}+c_{B,m_{i}})+{\rm H.c.}.
\label{interactionH}
\end{equation}
The giant atom system can be implemented on the mature superconducting quantum circuit platform, on which giant atom structures are constructed using capacitors, inductors, and Josephson junctions~\cite{BK2020,GA2019,AM2021}. On this platform, the frequencies of giant atoms can be finely tuned. Thus, we consider the transition frequency $\omega_{i}$ of the $i$th giant atom to be time-dependently tunable, while it always lies within the band gap. Here, $g_{i}$ is the coupling strength and $N$ denotes the number of giant atoms in the system.

We consider a system of two giant atoms coupled to the SSH chain, that is $N=2$.
In Fig.~\ref{energytwo}(a), we plot the energy spectrum of the system as a function of the frequency detuning $\Delta\Omega=\omega_{2}-\omega_{1}$ with $\theta=0.2\pi$.
It can observe that the system's energy spectrum can be divided into three parts. First, there is the energy band region, which includes two continuous energy bands where photons can propagate with corresponding group velocities.  Above and below the energy bands lie two bound state out of the continuum, arising from the giant atom breaking the translational symmetry of the SSH waveguide, as shown in the blue lines of Fig.~\ref{energytwo}(a). The final part consists of the band gap states located within the energy band gap, as shown by the red lines of the Fig.~\ref{energytwo}(a). The giant atoms act as effective boundaries within the system, inducing band gap states under periodic boundary conditions. As the eigenenergies of the band gap states lie within the bandgap, these states remain effectively decoupled from the bulk states and are prevented from dissipating into the bulk modes. Consequently, photon propagation is strongly suppressed within the band gap region.

In the model that a single giant atom is coupled to the SSH chain via $A-B$ coupling, the non-zero energy level within the band gap can be tuned on demand by adjusting the size of the giant atom~\cite{WC2022}, that is, the values of $n_{i}$ and $m_{i}$. Extending this property to the multi giant atoms system, we can achieve a more diverse band structure. As shown in Fig.~\ref{energytwo}(a), letting $d_{i}=|m_{i}-n_{i}|$, we find that in the two giant atom system, the two band-gap states exhibit crossing behavior when $d_{1}$ is even and $d_{2}$ is odd. In the single excitation subspace, the eigenwave function of the system can be expressed as
\begin{equation}
|\psi\rangle=\sum_{i=1}^{N}C_{i}^{e}\sigma_{i}^{+}|G,{\rm vac}\rangle+\sum_{l=1}^{L}(C_{A,l}c_{A,l}^{\dagger}+C_{B,l}c_{B,l}^{\dagger})|G,{\rm vac}\rangle.
\label{wavefunction}
\end{equation}
Here, $C_{i}^{e}$ is the giant atomic excitation amplitude and $C_{A(B),l}$ is the photon probability amplitudes for the $A(B)$ resonator. $|G\rangle$ is the ground state of the giant atom and the $|{\rm vac}\rangle$ is the vacuum state of the SSH chain. Through numerical calculations, we obtain the system's eigenvalues $E$ and their corresponding eigenwave functions $|\psi\rangle$. We select four eigenvalues with parameters $\Delta\Omega=0.65q$ (corresponding to points $a$ and $c$) and $\Delta\Omega=0.75q$ (corresponding to points $b$ and $d$) on both sides of the crossing point as shown in the black dots of the Fig.~\ref{energytwo}(a). To further verify this energy level crossing structure, we employ the fidelity method to compare the eigenwave function of the system. By comparing the fidelities of the corresponding eigenstates marked by black dots, we can further verify this crossing structure. The results yield $\mathcal{F}_{1} = |\langle\psi_{d}|\psi_{c}\rangle|^{2} = 0.0018$ and $\mathcal{F}_{2} = |\langle\psi_{b}|\psi_{c}\rangle|^{2}= 0.9982$. Here, the state on the left (right) side of the crossing point corresponds to the black dots of the Fig.~\ref{energytwo}(a), denoted by $|\psi_{x}\rangle (x=a,b,c,d)$. This indicates that after traversing the crossing point, the upper band gap state shifts toward the lower energy band, with analogous behavior observed for the lower band gap state.

Furthermore, the photonic spatial distributions $C_{l}^{c(d)}$ of the eigenstates $|\psi_{c}\rangle$ and  $|\psi_{d}\rangle$ are shown in Fig.~\ref{energytwo}(b,c). The results indicate that photons are tightly confined to the regions covered by the giant atoms. Specifically, on both sides of the crossing point, photons are localized at the left and right giant atoms, respectively. More notably, the photon distributions in Fig.~\ref{energytwo}(b) and (c) exhibit even more distinct behaviors. In Fig.~\ref{energytwo}(b), the photon distribution appears separated, while in Fig.~\ref{energytwo}(c), it is concentrated. To distinguish these two configurations for subsequent discussion, we refer to the one in Fig.~\ref{energytwo}(b) as the beam splitting state and that in Fig.~\ref{energytwo}(c) as the beam combining state.

\section{Manipulation of Photon Shape}
\label{evolution}

\begin{figure*}
\centering
\includegraphics[width=2\columnwidth]{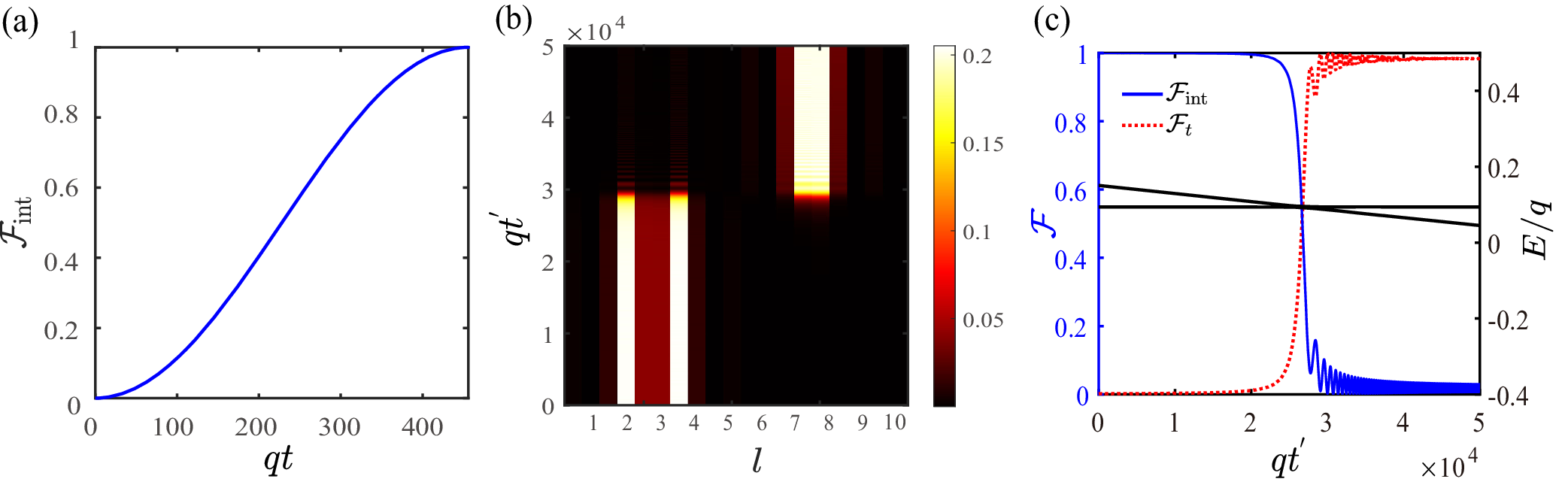}
\caption{(a) Time evolution of the fidelity $\mathcal{F}_{\rm int}$ during the driving process that defined as $\mathcal{F}_{\rm int}=|\langle\psi_{\rm int}|\psi(t)\rangle|^{2} $. (b) The photon probability distribution at the sites as a function of transfer time $t'$ during the transfer process. (c) The fidelity of the state transfer as a function of transfer time and energy spectrum of the system varies with transfer time. The blue solid line represents the fidelity with the initial state $\psi_{\rm int}$, and the red dashed line represents the fidelity with the target state $\psi_{t}$. The parameters set as $\theta=0.2\pi, L=10, n_{1}=2, m_{1}=4, n_{2}=7, m_{2}=8, \omega_{1}=0.4q, \Delta\Omega_{0}=0.6q, \delta\omega=0.2q, \delta=0.5q, \xi=0.005q$ and $g=0.9q$.}
\label{two}
\end{figure*}

In the previous section, we showed that a system of two giant atoms coupled to an SSH chain exhibits crossing structures within the band gap. Furthermore, we define combining state and splitting state based on the spatial shape of photon distributions. For the state transfer schemes in the topological waveguide, such as the Aubry-Andr\'{e}-Harper (AAH) model, have intrinsic channels connecting the edge states at both ends~\cite{YE2012}, allowing state transfer through adiabatic parameter variation. Inspired by this mechanism, we propose to manipulate photon states in our system by utilizing the analogous crossing structure. Our scheme is divided into two parts, namely the preparation process and the transfer process. The whole process is governed by the Schr\"{o}dinger equation, $-i\partial_{t}|\psi(t)\rangle=H(t)|\psi(t)\rangle$.

We begin with the preparation process, which aims to prepare the initial states for the state transfer. Initially, both giant atoms are in their ground states and the SSH chain is in the vacuum state. Moreover, during the preparation process the frequency of giant atoms remains constant. The driving field with frequency $\omega_{d}$ is applied to the first giant atom. The corresponding driving Hamiltonian given by
\begin{equation}
H_{d}=\Theta(t_{0}-t)\xi\sigma_{1}^{+}e^{-i\omega_{d}t}+{\rm H.c.},
\label{Hdrive}
\end{equation}
where $\xi$ represents the driving strength. $\Theta(t_{0}-t)$ is the Heaviside function, where $t_{0}$ denotes the driving duration for the first giant atom. In superconducting qubits system, the driving field can be implemented using independent circuits applied to the giant atoms~\cite{RB2014,PK2019}. The atomic frequencies  are set to the left of the crossing point, specifically with $\Delta\Omega_{0}=0.6q$. We select the beam splitting state $\psi_{s}$ as the initial state for the subsequent transfer process, that is $\psi_{\rm int}=\psi_{s}$. This state is prepared via resonant driving, with the drive frequency $\omega_d$ tuned to match the energy of the initial state, $\omega_d = E_s$. As shown in Fig.~\ref{two}(a), the fidelity between the system state and the initial state reaches a maximum value of $\mathcal{F}_{\rm int}=|\langle\psi(t)|\psi_{\rm int}\rangle|^{2}=1$ at $qt=445$, confirming the successful preparation of the initial state. Since this state lies within the band gap, it is protected from radiative decay into the continuum. Therefore, after turning off the driving field, the prepared initial state $\psi_{\rm int}$ remains stable.

We next proceed to the transfer process. At this stage, the frequency of the first giant atom remains constant, while the frequency of the second giant atom varies slowly as the function of the transfer time $t'$. Therefore, the interaction Hamiltonian can be further expressed as
\begin{eqnarray}
H_{I}(t')&=&\omega_1 |e\rangle_{11}\langle e|+\omega_2(t') |e\rangle_{22}\langle e|\nonumber\\
&&+\sum_{i}g[\sigma_{i}^{+}(c_{A,n_{i}}+c_{B,m_{i}})+{\rm H.c.}],
\end{eqnarray}
Here, the frequency of the second giant atom varies slowly with time, characterized by the frequency detuning $\Delta\Omega=\omega_1-\omega_2=\Delta\Omega_{0}+\delta\omega t'/T$. $\Delta\Omega_{0}$ is the atomic frequency detuning at the initial time of the transfer process, $T$ is the total transfer time, and $\delta\omega$ is the variation of the atomic frequency detuning after the transfer process.

We numerically solve the Schr\"{o}dinger equation over the entire evolution time $T$ to visualize the time evolution of the photon distribution. Fig.~\ref{two}(b) shows the evolution of the photon distribution as a function of the transfer time $t'$. The results indicate that the spatial location of the photon state transfers from the region covered by the left giant atom to that covered by the right giant atom. Meanwhile, the shape of the photon state also changes, transforming from the splitting state into the combining state. After a total transfer time $T$, the atomic frequency detuning is given by $\Delta\Omega=\Delta\Omega_{0}+\delta\omega=0.8q$. We define the combining state $\psi_{c}$ at this point as the target state, $\psi_{t}=\psi_{c}$. The fidelity $\mathcal{F}_{t}=|\langle\psi(t)|\psi_{t}\rangle|^{2}$, which measures the overlap between the system state and the target state, is plotted as a red dashed curve in Fig.~\ref{two}(c).
As shown in the figure, the fidelity initially remains at 0, then rapidly increases from 0 to nearly 1 around $qt'=2.66\times10^{4}$ reflecting the conversion of the beam splitting state into the beam combining state. Beyond this point, the fidelity remains constant. The blue curve in the same figure represents the fidelity between the system state and the initial state of the transfer process. It reveals that the system initially remained stable in its initial state, after which the fidelity rapidly dropped from 1 to 0. The evolution of the two relevant energy levels as a function of the transfer time $t'$ is depicted by the black lines in Fig.~\ref{two}(c). The transfer of the photon state from the beam splitting state to the beam combining state is clearly observed to occur at the crossing point.
In summary, we have demonstrated dynamic control of photonic states by utilizing the crossing structure and tuning the frequencies of the giant atoms. Specifically, the photon state transfers from the left to the right giant atom, while its spatial shape converts from the splitting state into the combining state. The positioning of the crossing structure inside the band gap guarantees high efficiency manipulation. This dynamic conversion from a splitting to a combining state aligns with the physical principles underlying key quantum communication architectures, such as quantum interferometers~\cite{RB2024,AO2016}, thereby extending the potential applications of giant atoms in this field.

\section{Shape-preserving Transfer}
\label{discussion}

\begin{figure}
\centering
\includegraphics[width=1\columnwidth]{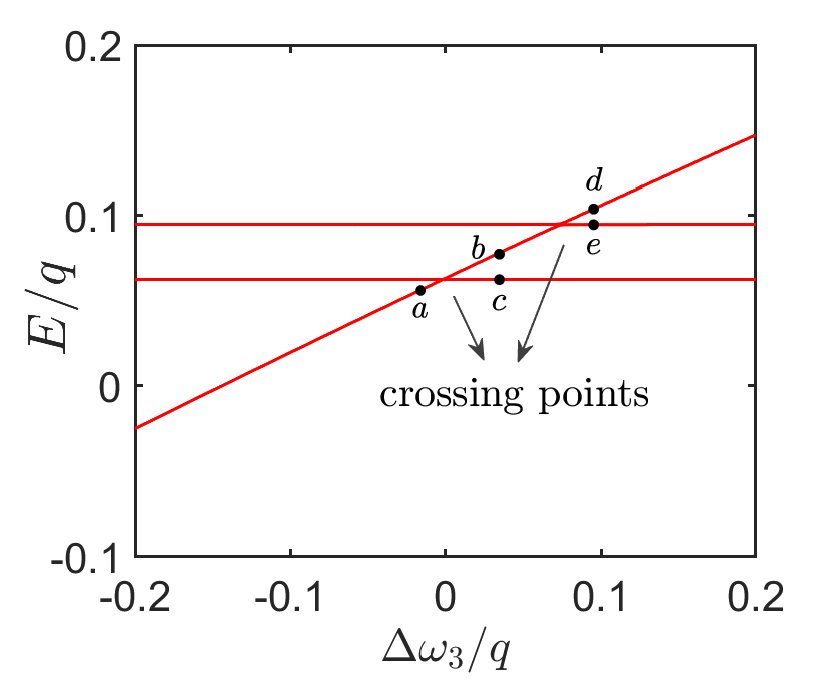}
\caption{The three crossing band gap states of the energy spectrum as a function of the frequency detuning in the three giant atoms system which the frequency detuning is $\Delta\omega_{3}=\omega_{3}-\omega_{1}$. The red solid lines denote the three crossing band gap states. The black points $a$ correspond to the frequency detuning $\Delta\omega_3=-0.02q$. The parameters selected for the black points $b$ and $c$ is $\Delta\omega_3=0.05q$. The frequency detuning of the black points $d$ and $e$ are $\Delta\omega_3=0.1q$. The parameter set as $\theta=0.2\pi, L=100, n_{1}=50, m_{1}=52, n_{2}=54, m_{2}=55, n_{3}=60, m_{3}=62, \omega_{1}=0.4q, \omega_{2}=-\omega_{1}=-0.4q, \delta=0.5q,$ and $g=0.9q$.}
\label{energythree}
\end{figure}

In the above section, we presented a dynamic control strategy for the shape of the photon state. Since photons serve as carriers for encoded information in quantum control processes, shape-preserving transfer of photon states is equally essential. To achieve this objective, we propose a scheme utilizing a system of three giant atoms coupled to an SSH waveguide.

We consider a system consisting of three giant atoms coupled to the SSH chain via $A-B$ coupling, whose interaction Hamiltonian can still be described by Eq.~(\ref{interactionH}). Following the idea of our proposed scheme, by adjusting the geometric configuration of the giant atoms, the crossing structures within the band gap can also be realized for the three giant atoms system with two crossing points formed as shown in Fig.~\ref{energythree}. Since our work focuses exclusively on the behavior of the band gap states, only these three states are presented. The crossing structure can be further verified by comparing the fidelities of the corresponding eigenstates marked by black dots. For the left crossing point, we obtain $\mathcal{F}_{1} = |\langle\psi_{b}|\psi_{a}\rangle|^{2} = 0.9986$ and $\mathcal{F}_{2} = |\langle\psi_{c}|\psi_{a}\rangle|^{2}= 0$. For the right crossing point, we have $\mathcal{F}_{3} = |\langle\psi_{d}|\psi_{b}\rangle|^{2} = 0.9938$ and $\mathcal{F}_{4} = |\langle\psi_{e}|\psi_{b}\rangle|^{2}= 0.0006$. Here, the states on the left (right) side of the crossing points correspond to the black dots in Fig.~\ref{energythree}, denoted as $|\psi_{x}\rangle (x=a,b,c,d,e)$. These results indicate that after traversing the two crossing points, the lowest band gap state shifts upward in energy.

\begin{figure}
\centering
\includegraphics[width=1\columnwidth]{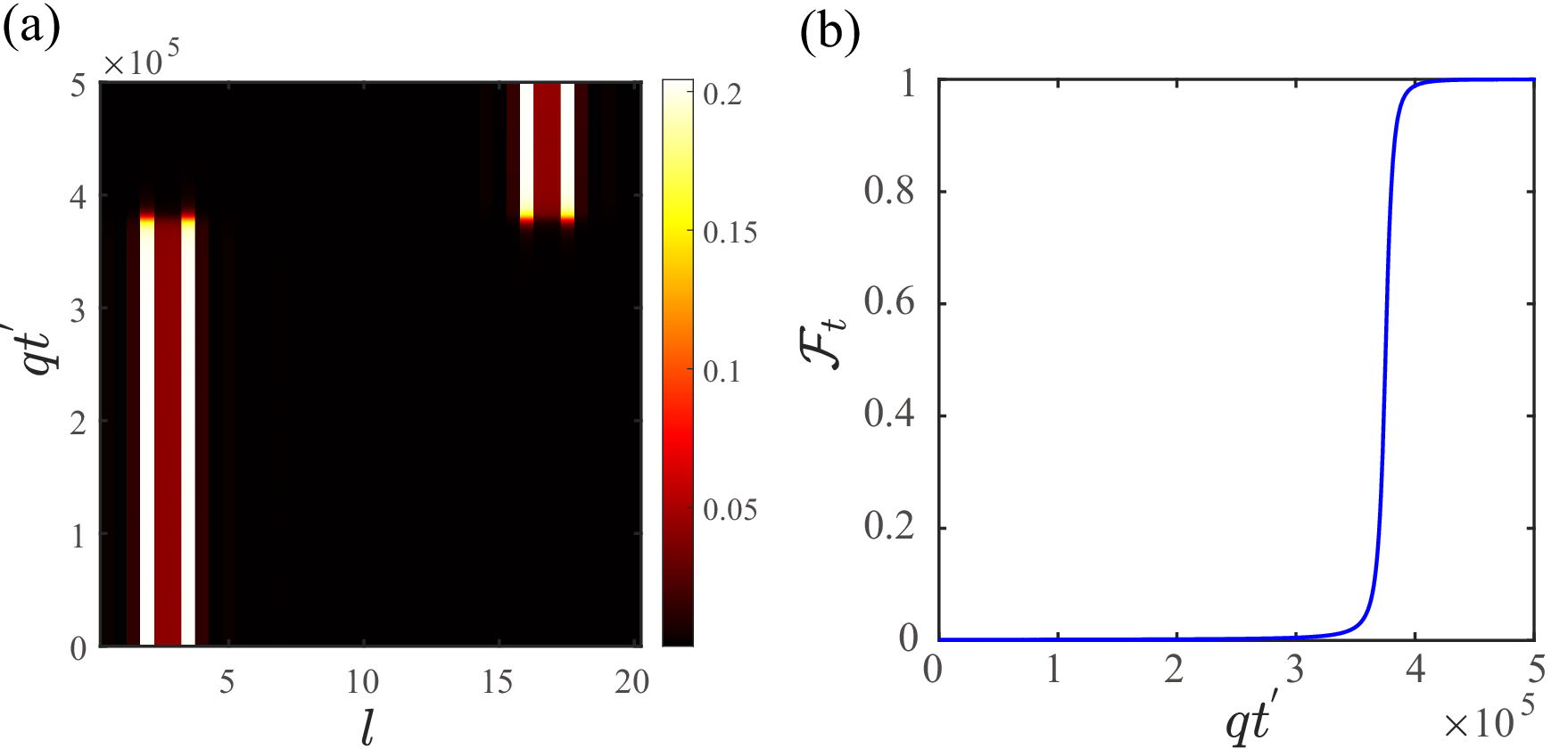}
\caption{(a) The photon probability distribution at the sites as a function of transfer time $t'$ during the transfer process for three giant atoms setup. (b) The fidelity of the state transfer as a function of transfer time. The blue solid line represents the fidelity with the target state $\psi_{t}$. The parameters set as $\theta=0.2\pi, L=20, n_{1}=2, m_{1}=4, n_{2}=7, m_{2}=8, n_{3}=16, m_{3}=18, \omega_{1}=0.4q, \omega_{2}=-0.4q, \Delta\omega_{3}(0)=-0.15q, \delta\omega_{3}=0.2q, \delta=0.5q, \xi=0.005q$ and $g=0.9q$.}
\label{three}
\end{figure}

Shape-preserving transfer scheme consists of two processes, which is the preparation process and the transfer process. The entire system dynamics is governed by the Schr\"{o}dinger equation.
We select the upper band gap state, which is the beam splitting state, as the initial state for the subsequent transfer process.
During the preparation stage, a driving field resonant with the initial state of the transfer process is applied to the left giant atom. The Hamiltonian of this drive field is still given by Eq.~(\ref{Hdrive}). Throughout the preparation process, the frequencies of all giant atoms are held fixed at $\omega_{1} = -\omega_{2} = 0.4q$ and $\Delta\omega_{3} = \omega_3 - \omega_1 = -0.15q$. Here, we set the length of the SSH waveguide as $L=20$. Once the desired initial state is successfully prepared, the driving field is turned off. The system then evolves freely under the Schr\"{o}dinger equation. In the subsequent transfer process, the frequency of the right giant atom is varied slowly as a function of the transfer time $t'$. The interaction Hamiltonian in the transfer process specified as
\begin{eqnarray}
H_{I}(t')&=&\omega_1 |e\rangle_{11}\langle e|+\omega_2 |e\rangle_{22}\langle e|+\omega_3(t') |e\rangle_{33}\langle e|\nonumber\\
&&+\sum_{i}g[\sigma_{i}^{+}(c_{A,n_{i}}+c_{B,m_{i}})+{\rm H.c.}].
\end{eqnarray}
Here, the frequencies of the left and middle giant atoms remain constant, while the frequency of the right giant atom varies slowly in time as $\omega_{3}(t')=\omega_{1}+\Delta\omega_{3}(t')$. The frequency detuning follows $ \Delta\omega_{3}(t')=\Delta\omega_{3}(0)+\delta\omega_{3} t'/T$, with $\Delta\omega_{3}(0) =-0.15q$. As illustrated in Fig.~\ref{three}(a), the initial state is a splitting state where photons are distributed in the region covered by the left giant atom. After remaining stable for a period of time, they are transferred from the left giant atom to the right giant atom near the second crossing point. The fidelity with respect to the target photonic splitting state at the right atom reaches $100\%$, as numerically calculated and shown in Fig.~\ref{three}(b). This result confirms the realization of shape-preserving transfer in our proposed scheme.

\section{Conclusion}
\label{conclusion}
In conclusion, we have investigated the system of giant atoms coupled to an SSH chain waveguide. We show that by designing the geometric configuration of the atoms, an energy-level crossing can be engineered within the topological band gap. Leveraging its unique band structure, we achieve dynamic control over the photonic state shape by tuning the frequencies of the giant atoms. The efficiency of this control scheme is enhanced by the location of the level crossing inside the gap. Specifically, for two giant atoms system, we demonstrate that such an energy level crossing structure allows dynamic modulation of the photonic state shape. By tuning the atomic frequencies, the shape of the photon state evolves from a splitting state into a combining state as it transfers from the left to the right giant atom. Extending the scheme to three giant atoms, we further demonstrate the feasibility of shape-preserving photon state transfer.

On superconducting quantum circuit platforms, giant atoms can be realized with transmon qubits~\cite{BK2020,GA2019,AM2021}, and coupled resonator waveguides of tens of sites have been fabricated~\cite{PR2017,XYZ2023}. Building on this platform, the slow tuning of atomic frequencies has become experimentally feasible. By employing enlarged capacitive couplings in lumped-element designs, nearest-neighbor hopping strengths on the order of MHz have been achieved. In currently feasible experiments, the lifetime of artificial atoms composed of superconducting qubits is on the order of $10\mu s$~\cite{MK2020}. From the above comparison, our scheme demonstrates a feasible approach that combines the self-interference effect of giant atoms with the topological properties of the SSH chain, offering a promising platform for quantum information processing.

\begin{acknowledgments}

Z.W. is supported by Natural Science Foundation of China (Grants Nos. 12375010), the Quantum Science and Technology-National Science and Technology Major Project (No. 2023ZD0300700).

\end{acknowledgments}

\end{document}